# Band gaps and structural properties of graphene halides and their derivates: A hybrid functional study with localized orbital basis sets

František Karlický,[a)] Radek Zbořil, Michal Otyepka [b)]

*Regional Centre of Advanced Technologies and Materials, Department of Physical Chemistry, Faculty of Science, Palacký University, Tř. 17. listopadu 12, Olomouc 771 46, Czech Republic*

DFT calculations of the electronic structure of graphane and stoichiometrically halogenated graphene derivatives (fluorographene and other analogous graphene halides) show (i) localized orbital basis sets can be successfully and effectively used for such 2D materials; (ii) several functionals predict that the band gap of graphane is greater than that of fluorographene, whereas HSE06 gives the opposite trend; (iii) HSE06 functional predicts quite good values of band gaps w.r.t benchmark theoretical and experimental data; (iv) the zero band gap of graphene is opened by hydrogenation and halogenation and strongly depends on the chemical composition of mixed graphene halides; (v) the stability of graphene halides decreases sharply with increasing size of the halogen atom - fluorographene is stable, whereas graphene iodide spontaneously decomposes. In terms of band gap and stability, the $C_2FBr$, and $C_2HBr$ derivatives seem to be promising materials, e.g., for (opto)electronics applications, because their band gaps are similar to those of conventional semiconductors, and they are expected to be stable under ambient conditions. The results indicate that other fluorinated compounds ($C_aH_bF_c$ and $C_aF_bY_c$, Y = Cl, Br, I) are stable insulators.

---

[a)] Electronic mail: frantisek.karlicky@upol.cz

[b)] Electronic mail: michal.otyepka@upol.cz

## I. INTRODUCTION

Graphene, a two-dimensional material first prepared in 2004, has remarkable mechanical, electrical and optical properties.[1,2] It is regarded as one of the most promising candidates for the next generation of electronic materials due to its extremely high charge carrier mobility. However, graphene lacks a band gap around the Fermi level, which is a defining feature of semiconductor materials and essential for controlling conductivity by electronic means.[3] This greatly restricts its uses in electronics, and therefore various ways to generate tunable gaps in the energy spectrum of graphene have been suggested and explored. Notably, it has been shown that the electric conductivity of pure graphene can be modified by chemical doping, adding impurities, noncovalent modification, and chemical functionalization. In the present study, we examined stoichiometrically hydrogenated and halogenated graphene derivatives, which are of interest to both experimentalists and theoreticians due to their broad range of potential applications. Several new graphene-based two-dimensional (2D) crystals, i.e., fully hydrogenated graphene (graphane, CH)[4,5] and fully fluorinated graphene (fluorographene, CF, or graphene fluoride)[6-8], have been prepared recently.

In general, time-independent density functional theory (DFT) is used in most calculations of the electronic structures of solid-state materials. The main limitation of the DFT approach is it is inherently a ground-state theory. Local density approximation (LDA) and generalized gradient approximation (GGA) functionals systematically underestimate Kohn-Sham band gaps (compared to experimentally determined values), whereas the Hartree-Fock method systematically overestimates them.[9] Hybrid functionals often give reasonably accurate predictions of band gaps, but these functionals are computationally demanding due to the slow decay of Hartree-Fock exchange.[10] In addition, band gaps calculated using these methods are often highly sensitive to the identity of the functional used and sometimes produce results that are inconsistent with experimental data or results obtained using more computationally demanding methods such as GW, Bethe-Salpeter equation (BSE) or quantum Monte Carlo;[9] for example, the band gap of pyrite calculated using B3LYP has been shown to be twice as large as the experimental value.[11] Short-range functionals, such as the screened hybrid functional of Heyd, Scuseria, and Ernzerhof (HSE),[10] seem to be an effective alternative to standard hybrid functionals. The HSE functional accurately predicts the electronic properties of low-dimensional carbon materials and optical transitions in both metallic and semiconducting single-wall carbon nanotubes.[12] We therefore expected that HSE functional calculations would be useful for accurately describing the electronic structure of the systems considered in our study.

Of the materials considered in this study, benchmark data based on high-level calculations are only available for graphane and fluorographene (graphene fluoride). Experimental measurements have determined that graphane is an insulator.[4,5] However, the exact band-gap value has not yet been published; the only value reported was the lower bound of the band gap for single-side hydrogenated graphene of 0.5 eV.[4] Conventional DFT GGA calculations have revealed that graphane is a direct band-gap material and estimated that the band gap is about 3.5 eV at the Γ point (see Table I).[13-17] However, GW calculations have suggested band-gap values of 5.4 eV,[14,18] 5.7 eV[17] and 6.1 eV.[15,19] Recent high-level theoretical calculations of optical properties carried out by BSE and accounting for electron-electron and electron-hole correlations and excitonic effects have shown that the first exciton peak in the (in-plane) absorption spectrum of graphane occurs at 3.8 eV.[18]

TABLE I. Literature values of band gaps (eV) calculated for chair conformations of graphane (CH) and fluorographene (CF).[a]

| Method | Basis | Code | $E_g$ | Source |
|---|---|---|---|---|
| *Graphane (CH)*[b] | | | | |
| DFT(GGA) | PW | CASTEP | 3.5 | Ref. [20] |
| DFT(LDA) | PW | Abinit | 3.6 | Ref. [21] |
| DFT(GGA) | NAO | Siesta | 3.8 | Ref. [13] |
| DFT(GGA) | PW | VASP | 3.5 | Ref. [14] |
| GW over LDA | PW | VASP | 5.4 | Ref. [14] |
| DFT(LDA) | PW | VASP | 3.4 | Ref. [22] |
| DFT(GGA) | PW | Quantum Espresso | 3.5 | Ref. [17] |
| GW over GGA | PW | Quantum Espresso | 5.7 | Ref. [17] |
| DFT(GGA) | PW | Abinit | 3.7 | Ref. [15] |
| GW over GGA | PW | Yambo | 6.1 | Ref. [15] |
| DFT(GGA) | PW | VASP | 3.5 | Ref. [16] |
| DFT(LDA) | PW | VASP | 3.4 | Ref. [19] |
| GW over LDA | PW | VASP | 6.0 | Ref. [19] |
| GW over LDA | PW | Yambo | 5.4 | Ref. [18] |
| BSE over GW-LDA[c] | PW | Yambo | 3.8 | Ref. [18] |
| *Fluorographene (CF)* | | | | |
| DFT(LDA)[d] | PW | Corning | 3.5 | Ref. [23] |
| DFT(LDA)[d] | PW | FHI | 3.0 | Ref. [24] |
| DFT(GGA) | NAO | Siesta | 4.2 | Ref. [13] |
| DFT(LDA) | PW | VASP | 3.0 | Ref. [22] |
| DFT(GGA) | PW | VASP | 3.1 | Ref. [16] |
| DFT(GGA) | PW | Quantum Espresso | 3.0 | Ref. [7] |
| DFT(GGA) | PW | VASP | 3.1 | Ref. [25] |
| DFT(GGA) | PW | Abinit | 3.2 | Ref. [15] |
| GW over GGA | PW | Yambo | 7.4 | Ref. [15] |
| GW over LDA | PW | VASP | 7.5 | Ref. [19] |
| GW over LDA | | | | |
| GW over GGA | PW | Yambo | 7.5 | Ref. [26] |
| BSE over GW-GGA[c] | PW | Yambo | 5.4 | Ref. [26] |
| Exp. | --- | --- | 3.0 | Ref. [6] |
| Exp. | --- | --- | 3.8 | Ref. [27] |

[a] Abbreviations: plane waves (PW), numerical atomic orbitals (NAO). [b] No experimental data available; only lower bound of band gap of 0.5 eV for single-sided hydrogenated graphene is reported in Ref. 4 [c] First exciton peak of in-plane optical spectrum. [d] Calculation for graphite monofluoride.

Reported band-gap values for fluorographene are also controversial (see Table I). Experimental measurements have shown that fluorographene is an insulator with a band gap of 3.0 eV (extracted from optical spectra) and resistivity higher than $10^{12}$ Ω.[6,28] Photoluminescence measurements have identified an emission peak at 3.80 eV in the fluorographene dispersion spectrum in acetone, which has been assigned to band-to-band recombination of a free electron and hole.[27] Theoretical calculations of the fluorographene band gap by GGA DFT have suggested values around 3.1 eV,[7,15,16,22,25] similar to values for graphite fluoride.[23,24] The agreement between theoretical and experimental band-gap values is probably accidental because high-level theoretical calculations using GW predict the band gap to be around 7.5 eV.[15,19,26] Finally, calculation of optical spectra by BSE on top of GW predicted an spectral onset at around 5.4 eV.[26] This value was considered to be in good agreement with experimental values (3.0 and 3.8 eV from optical measurements), taking into account effects of corrugation and defects.[26]

In the work described in this paper, we examined computational approaches for such class of 2D materials based on DFT and localized orbital basis sets and compared the calculated results with available benchmark data. Besides the popular PBE, we considered six additional functionals, including the promising screened hybrid functional HSE06. Most of our calculations were carried out with localized orbital basis sets for two main reasons. Firstly, it meant that a rather "small" number of localized orbitals was required to achieve converged results in 2D (see Sec. III.A), which in turn led to lower computational demands. The second reason was essentially practical as packages based on localized orbitals offer a broader set of functionals with respect to plane-wave (PW) codes. Recent theoretical and experimental data (collected in Table I) were used to identify functionals which could accurately predict the properties of the 2D materials considered (Sec III.B). The results showed that the band gaps of graphene halides depend on their stoichiometry and can be relatively finely tuned between zero and the band gap of graphane or fluorographene. We also estimated the stability of these materials to assess their potential utility in empirical investigations (Sec. III.C).

## II. COMPUTATIONAL DETAILS

The species studied were halogenated graphane analogues with empirical formula $C_aX_bY_c$ ($a = b + c$; X = H, F; Y = F, Cl, Br, I; Y≠X). In all cases, the species were initially assumed to adopt a chair-like conformation because it has been shown to be the most stable conformation of both graphane (CH) and fluorographene (CF).[14,20] The model consisted of an infinite 2D structure with the smallest possible supercell under periodic boundary conditions (PBC). Infinite structures were modeled using linear-scaling DFT with Gaussian orbitals and PBC.[29] Such calculations were performed using GAUSSIAN 09[30], which is an effective tool for applying hybrid functionals, since the Gaussian orbital used makes a hybrid functional application very efficient and convenient.[11] For comparison, calculations on the simplest systems were also performed with plane-wave basis sets; the Vienna ab initio simulation package (VASP)[31] implementing the projector augmented waves (PAW) method[32] was used. The LDA, PBE, BPW91, BLYP, TPSS, M06L, and HSE06 functionals were tested in conjunction with large basis sets (Sec. III.A), which generate almost equivalent band gaps and geometries as plane-wave basis sets and effective core potentials for heavy elements were included. Structures corresponding to energy minima were obtained by optimizing all coordinates and unit cell lengths using the default convergence criteria in GAUSSIAN. In VASP, the optimized unit cell has been obtained minimizing the total energy as a function of the lattice parameter. At each value of the lattice constant the atomic positions (i.e. the internal degree of freedom) were fully relaxed. The total energies calculated for optimal structures were used to evaluate the stability of the species relative to graphane by calculating a difference $\Delta E´$ of total energies according to Eq. 1. A *k*-point mesh of at least 16×16 points was used to sample the Brillouin zone of the smallest supercells.

## III. RESULTS AND DISCUSSION

After identification of suitable basis set for several systems, we employed different functionals to obtain band gaps and structural parameters of graphane and fluorographene. Finally, as a case study, we calculated band gaps and stabilities of mixed graphene halides.

### A. Basis set choice

There are, in principle, two types of localized orbital basis sets: basis sets designed for solids (e.g., used in Crystal code[33]) or molecular basis sets. All basis sets used in this work were molecular Gaussian basis sets (except for the comparative calculations with plane-wave basis sets). Scuseria and coworkers[34] have shown that the triple-$\zeta$ basis set is sufficient for modeling both structural and electronic properties of almost all elements in 3D materials, but the lightest elements were not considered. We performed extensive tests on several systems to identify basis sets that could accurately predict structural parameters and band gaps for the class of 2D materials considered. We repeated convergence tests for PBE, BLYP and HSE06 functionals. As an example, Table II shows the basis-set dependence of the structural parameters and band gap of graphane, for which the dependence was strongest.

TABLE II. Band gap $E_g$ (eV) and geometrical parameters (Å and deg) for graphane as a function of basis-set size. The calculations were carried out with PBE (left subcolumn) and HSE06 (right subcolumn) functionals.

| Basis set for C/H | $E_g$ PBE | $E_g$ HSE | $d$(C-C) PBE | $d$(C-C) HSE | $d$(C-H) PBE | $d$(C-H) HSE | $d$(H-H) PBE | $d$(H-H) HSE | $a$(C-C-C) PBE | $a$(C-C-C) HSE |
|---|---|---|---|---|---|---|---|---|---|---|
| VDZ | 5.5 | 6.8 | 1.55 | 1.54 | 1.12 | 1.11 | 2.56 | 2.54 | 111.31 | 111.26 |
| 6-31G** | 5.0 | 6.2 | 1.54 | 1.53 | 1.11 | 1.11 | 2.55 | 2.53 | 111.50 | 111.46 |
| cc-pVDZ | 4.5 | 5.6 | 1.54 | 1.53 | 1.12 | 1.11 | 2.55 | 2.53 | 111.53 | 111.52 |
| 6-311G** | 4.3 | 5.4 | 1.54 | 1.53 | 1.11 | 1.10 | 2.54 | 2.53 | 111.45 | 111.41 |
| cc-pVDZ/cc-pVTZ | 4.2 | 5.3 | 1.54 | 1.53 | 1.11 | 1.10 | 2.55 | 2.53 | 111.46 | 111.41 |
| cc-pVTZ | 4.1 | 5.2 | 1.54 | 1.53 | 1.11 | 1.10 | 2.54 | 2.52 | 111.50 | 111.47 |
| cc-pVTZ/cc-pVQZ | 3.9 | 4.9 | 1.54 | 1.53 | 1.11 | 1.10 | 2.54 | 2.52 | 111.50 | 111.47 |
| cc-pVTZ/cc-pV5Z | 3.7 | 4.7 | 1.54 | 1.53 | 1.11 | 1.10 | 2.54 | 2.52 | 111.50 | 111.47 |
| cc-pVTZ/cc-pV6Z | 3.6 | 4.6 | 1.54 | 1.53 | 1.11 | 1.10 | 2.54 | 2.52 | 111.50 | 111.47 |
| PW ($E_{cut}$= 500 eV) | 3.5 | 4.5 | 1.54 | 1.53 | 1.10 | 1.10 | 2.54 | 2.52 | 111.63 | 111.54 |

The results show that the geometry was not very sensitive to basis-set choice and in general, a double-ζ type basis set was adequate. In contrast, the band gap was found to be strongly dependent on basis set. Surprisingly, the hydrogen basis set needed to be very large, up to at least a sextuple-ζ basis set, to achieve converged results (Table II). Our final choice of basis sets for optimization/single-point calculations was 6-311G**/cc-pV6Z for H and 6-311G*/cc-pVTZ for C, F, and Cl. Geometries and band gaps of 2D materials were sufficiently well converged with such basis sets. It should be noted that for heavy elements, like Br and I, relativistic effects come into play. Therefore, we included them through the relativistic effective core potential (RECP). Large-core RECPs (we tested LANL2DZ) were not so suitable because lattice constants were overestimated by up to 0.1 Å for the largest supercells and predicted band gaps were underestimated by up to 0.5 eV with respect to small-core RECPs; however, trends were preserved. Sufficient basis sets in combination with small-core RECP for Br and I were cc-pVDZ-PP and cc-pVTZ-PP for optimization and additional single-point calculations, respectively, in agreement with a previous study on 3D materials.[34]

For the simplest systems (CX, X = H, F, Cl, Br), we also compared results obtained with Gaussian basis sets to those with plane-wave basis sets, where the size of the basis set is determined by the cut-off energy parameter $E_{cut}$. The predicted geometrical parameters were almost identical for all the CX structures (for CH *cf.* Table II). Band gaps calculated using a localized basis set were less than for plane-wave basis sets by up to 0.1 eV (*cf.* the last lines of Table II), which may indicate that the large localized basis sets still were not of sufficient size. It should be noted that the performance of the Gaussian calculations was very good, particularly when the hybrid functional was used; the code based on localized orbitals ran faster than the plane-wave code. Nonetheless, it should be pointed out that the Gaussian calculations were all-electron or small-core RESP calculations, whereas the VASP calculations were full-potential valence-only calculations. This means the calculations with local basis sets were particularly computationally expensive for molecules containing heavy elements. In addition, the default convergence criteria in Gaussian are stricter than in VASP. Thus, from a cpu-time viewpoint very effective approach is to use localized orbital basis sets in 2D regime, especially if hybrid functionals are needed.

## B. Calculations using different functionals

Due to the controversial differences between GGA PBE functional predictions and high-level GW and BSE results for graphane and fluorographene (Sec. I and Table I), several functionals, including the perspective hybrid short-range HSE06 functional, were tested (Table III). Full optimization of structures and additional single-point calculations with large basis sets were performed (see preceding Sec. III.A).

TABLE III. Band gaps (in eV) and geometrical parameters (Å and deg.) for graphane and fluorographene and their dependence on density functional.

| functional | $E_g$ | $d$(C-C) | $d$(C-X) | $d$(X-X) ≡ TV | $a$(C-C-C) |
|---|---|---|---|---|---|
| *Graphane (CH)* | | | | | |
| LDA   | 3.5 | 1.52 | 1.12 | 2.50 | 111.55 |
| PBE   | 3.6 | 1.54 | 1.11 | 2.54 | 111.49 |
| BPW91 | 3.9 | 1.54 | 1.11 | 2.54 | 111.47 |
| BLYP  | 3.7 | 1.55 | 1.11 | 2.56 | 111.50 |
| TPSS  | 3.8 | 1.54 | 1.10 | 2.54 | 111.32 |
| M06L  | 4.7 | 1.53 | 1.10 | 2.52 | 111.46 |
| HSE06 | 4.6 | 1.53 | 1.10 | 2.52 | 111.45 |
| *Fluorographene (CF)* | | | | | |
| LDA   | 3.1 | 1.55 | 1.36 | 2.55 | 110.33 |
| PBE   | 3.3 | 1.58 | 1.38 | 2.60 | 110.68 |
| BPW91 | 3.3 | 1.58 | 1.38 | 2.61 | 110.75 |
| BLYP  | 3.5 | 1.59 | 1.39 | 2.62 | 110.88 |
| TPSS  | 3.5 | 1.58 | 1.38 | 2.60 | 110.85 |
| M06L  | 4.1 | 1.56 | 1.37 | 2.58 | 110.76 |
| HSE03 | 5.2 | 1.56 | 1.36 | 2.58 | 110.65 |
| HSE06 | 5.1 | 1.57 | 1.36 | 2.58 | 110.63 |

The results show that the geometry is not very sensitive to functional choice. Maximal differences of 0.02 (0.04) Å, 0.02 (0.03) Å, and 0.06 (0.07) Å for C-C distance, C-H (C-F) distance and translation vector, respectively, were obtained for graphane (fluorographene). All tested functionals predicted larger band gaps for graphane than for fluorographene (difference 0.4 eV for LDA, 0.3 PBE, 0.6 BPW91, 0.2 BLYP, 0.3 TPSS, and 0.6 M06L), with the exception of the HSE06 functional (-0.5 eV). The HSE06 functional predicted a trend for CH and CF band gaps in qualitative agreement with benchmark results (see Table IV). The results indicate that the inclusion of some portion of exact Hartree-Fock exchange was crucial not only for increasing of predicted band-gaps but also for correct trend. In addition, HSE06 values of the band gap (4.5 eV and 5.1 eV for CH and CF) were quantitatively quite good, especially with respect to the high-level BSE results (3.8 eV and 5.4 eV). In the case of fluorographene, the HSE06 band gap of 5.1 eV is in agreement with the experimental estimate of the optical gap of ~ 3 eV[6] or 3.8 eV[27] after taking effects of corrugation and defects into account.[15,26] The plane-wave HSE06 calculations gave almost same band gap as calculated with localized orbitals. Comparison of the PBE and HSE06 band structures near the band gaps and corresponding densities of states (DOS) are shown in Figure 1 for both graphane and fluorographene. Surprisingly, the HSE06 band structures were almost the same as for PBE, except the HSE06 structures were more expanded. Based on these results, we employed the HSE06 functional in subsequent calculations of the studied 2D graphene-based materials.

TABLE IV. Summary and comparison of benchmark and calculated band gaps (in eV) for graphane and fluorographene. For details, see Table I.

| method / material | CH | CF |
|---|---|---|
| DFT(PBE)[a] | 3.5 | 3.1 |
| DFT(HSE06)[a] | 4.5 | 5.1 |
| BSE (spectra) | 3.8 | 5.4 |
| GW | 5.4-6.1 | 7.4-7.5 |
| Exp. | --- | 3.0, 3.8 |

[a] This work.

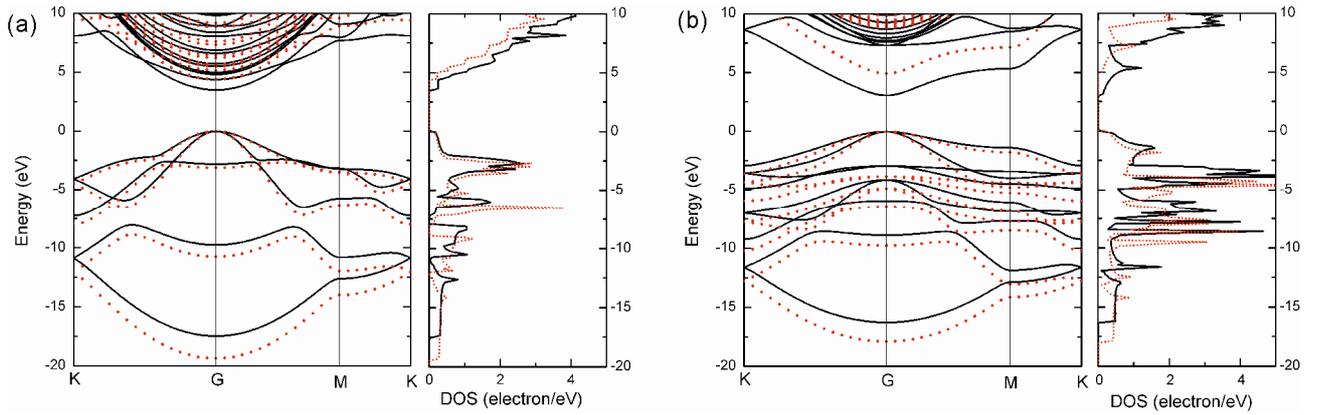

FIG. 1. The electronic band structure and projected density of states in the vicinity of the band gap for graphane (a) and fluorographene (b) along lines connecting the high symmetry points K, Γ, and M in the Brillouin zone. Band structures and DOS are calculated using the PBE (full black line) and HSE06 approximations (dotted red line). The Fermi level is set at zero energy.

### C. Band gaps and stability of graphene halides

Graphene exhibits metallic behavior at a single point in reciprocal space, the K-point, where the conduction and valence bands touch.[1,2] The hydrogenation or halogenations of graphene creates a finite band gap, transforming the graphene into a semiconductor or insulator. All of the optimized $C_aX_bY_c$ (X = H, F; Y = F, Cl, Br, I for Y≠X) compounds considered here are direct band-gap materials. The bottom of the conduction-band and top of the valence-band are located at the Γ point in the first Brillouin zone for all $C_aX_bY_c$ species. The top of the valence band is doubly degenerate and the maximum band gaps are located at the K points in CH, CF, CCl and CBr. The minimal direct HSE06 band gaps, $E_g$, at the Γ point are shown in Figure 2a for all $C_aX_bY_c$ compounds considered.

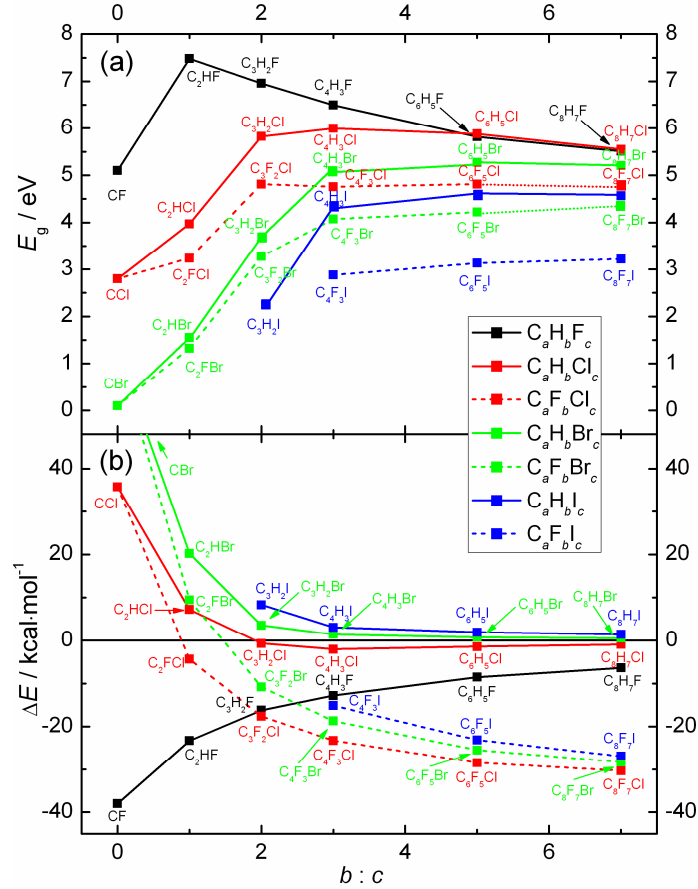

FIG. 2. Electronic band gaps (a) and stabilities (b, calculated using Eq. 1) of graphene derivates $C_aX_bY_c$, ($a = b + c$) as a function of the stoichiometric ratio $b$:$c$ for selected substituents X = H, F and Y = F, Cl, Br, I calculated using the HSE06 functional. The limits of $b$:$c \to 0$ [$\infty$] correspond to CF, CCl, and CBr [CH and CF].

Figure 2a shows how the band gap depends on the stoichiometry (chemical composition) of the graphene halides, suggesting it can be relatively smoothly tuned from the maximal value in graphane to almost zero in graphene bromide. Even finer tuning could in principle be achieved by adjusting the ratio of more than two substituents. The band gaps calculated using HSE06 are in quite good agreement with the reference data (Sec. III.B). Assuming that the predicted HSE06 band gaps for species without available reference data are also accurate, most of the considered species appear to have band gaps typical of insulators. Two species have band gaps comparable to those of conventional semiconductors. Specifically, the band gaps of $C_2FBr$ and $C_2HBr$ are 1.32 and 1.55 eV, respectively (*cf.* values of traditional semiconductors Si, Ge, and GaAs are 1.2, 0.7, and 1.5 eV, respectively).

We have previously shown that some graphene halides are unstable under ambient conditions (e.g., graphene iodide[8]). Further, all attempts at optimizing the geometry of graphene iodide (and some other hydrogenated and fluorinated graphene iodides) resulted in its decomposition to graphene and molecular iodine, which is consistent with experimental observations.[8] It was therefore considered necessary to assess the stability of the considered graphene derivatives; this was carried out (for single supercells) using Eq. 1:

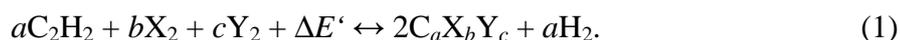

$$a\text{C}_2\text{H}_2 + b\text{X}_2 + c\text{Y}_2 + \Delta E' \leftrightarrow 2\text{C}_a\text{X}_b\text{Y}_c + a\text{H}_2. \quad (1)$$

Here, the product $C_aX_bY_c$ ($a = b + c$; X = H, F; Y = F, Cl, Br, I for Y≠X) is the target material and $\Delta E'$ is the difference between the sums of the total energies of the products and the reactants. To compare the stability of different materials, we considered the normalized energy difference $\Delta E = \Delta E'/(2a)$, where $a$ is the number of C atoms in a computational supercell. The choice of a reference material assigned with zero $\Delta E$ (and reactant in Eq. 1) was arbitrary; nonetheless, we preferred stable and experimentally prepared material. Of two such available 2D graphene-based materials, graphane (CH) and fluorographene (CF), we chose the former one with smaller stability since we obtained simple indication of the stability from the $\Delta E$ value. If a $C_aX_{a-c}Y_c$ compound is more stable than CH, $\Delta E$ is negative. If $\Delta E$ is positive, the species is less stable than CH but not necessarily thermodynamically unstable. The lower bound of stability was considered to be $\Delta E$ for CCl because it is predicted to be unstable under ambient conditions (the pristine parent material, graphite chloride, is unstable at temperatures > 0° C but stable at lower temperatures[35]). This assumption was supported by the fact that nonstoichiometric graphene chloride with low concentrations of approximately 8 at. % Cl has recently been prepared by photochemical chlorination of graphene.[36] In addition, very recently, preparation of few-layer graphene chlorinated up to 30 at. % and brominated up to 4.8 at. % by UV irradiation in liquid-chlorine/bromine medium has been reported.[37] Such chlorinated samples were shown to be stable at room temperature and the chlorine can be removed by heating to around 500 °C or laser irradiation. The Δ$E$ values (Figure 1b) indicate that the most stable of the graphene halide derivatives is fluorographene (CF), which has been prepared recently.[6-8] The stabilities predicted using HSE06 were in good agreement with those obtained using BLYP, which we performed for comparison. In contrast to the electronic properties, the thermodynamic properties of the considered materials were described sufficiently well using the BLYP functional.

The decrease in stability on going from CF to CCl to CBr and finally to CI (which spontaneously decomposes to iodine and graphene[8]) is probably a consequence of the halogen atoms being forced into excessively close proximity by the graphene halide scaffold. That is to say, the halogen-halogen distances in the CX species (CF: 2.58 Å; CCl: 2.89 Å; CBr: 3.06 Å) are less than the sum of the two halogens' van der Waals radii (F: 1.47 Å; Cl: 1.75 Å; Br: 1.85 Å). There is thus non-negligible overlap between the heavy halogen atoms, which leads to lengthening of the C-C bond, weakening of the C-X bond and destabilization of the graphene halide (see Figure 3).

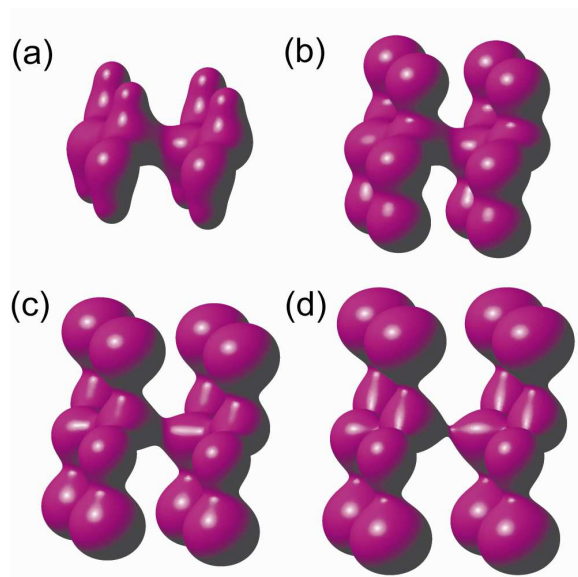

FIG 3. Total electron densities for graphane (CH) (a), fluorographene (CF) (b), graphene chloride (CCl) (c), and graphene bromide (CBr) (d) in a 2×2 supercell for isovalues of 0.1 a.u. The distances shown correspond to optimized geometries; the C-C distances are 1.53 Å (a), 1.57 Å (b), 1.74 Å (c), and 1.84 Å (d). Notice decreasing electron density in the middle of the C-X bond and the C-C bond with increasing size of the halogen atom.

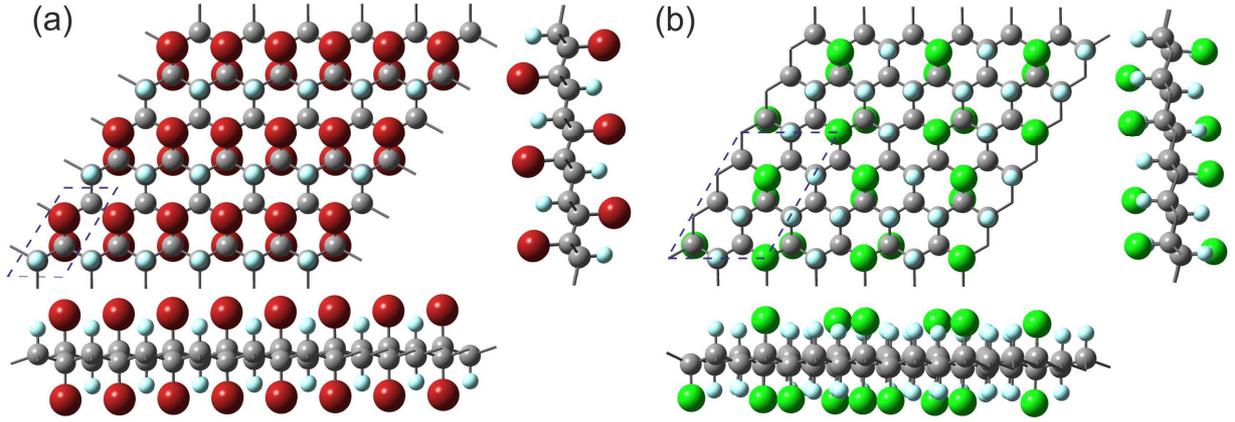

FIG 4. Optimized geometries of selected compounds whose stability is expected: (a) promising semiconductor $C_2FBr$, (b) insulator $C_3F_2Cl$. Carbon, fluorine, chlorine, and bromine atoms are shown as black, light blue, green, and red spheres, respectively. Computational supercells are designated by dotted line.

Two of the graphene halides examined in this work ($C_2FBr$ and $C_2HBr$) exhibit both semiconductor-like band gaps and stabilities comparable to CH or greater than that of CCl (see Figure 2), and thus stand out as promising materials for practical applications. In contrast, hydrogenated graphene fluorides ($C_aH_bF_c$) are generally more stable than graphane and their band gaps are very wide, while other hydrogenated graphene halides ($C_aH_bY_c$) are less stable than graphane (see Figure 2). The stable fluorinated graphene halides ($C_aF_bY_c$) are the most promising insulators, exhibiting good stability and band gaps that can be tuned to the semiconducting $C_2FBr$. For illustrative purposes, the optimized geometries of the semiconducting $C_2FBr$ and insulating $C_3F_2Cl$ are shown in Figure 4.

## IV. CONCLUSIONS

The electronic structure, namely band gap, of graphene halides depends strongly on their composition. The band gaps of the studied graphene halides cover a range from 0 to 7.5 eV. From a computational viewpoint, the use of localized orbital basis sets for such 2D materials is an efficient approach, especially if hybrid functionals are needed. GGA functionals systematically underestimate the band gaps of 2D carbon-based materials (graphene derivatives) with respect to available benchmark data, whereas the screened hybrid functional HSE06 provides band gaps that agree well with these reference values. It should be noted that the available reference data are currently very limited. On the other hand, screened hybrid functionals have been shown to perform well on a wide range of 3D materials[34] and some low-dimensional carbon materials.[12] While the calculated absolute band-gap values are sensitive to the choice of functional and the trends in band gaps are frequently opposite for CH and CF, the predicted stability of the materials is relatively insensitive. Several graphene halide derivatives whose calculated band gaps and stabilities would make them potentially suitable for use in semiconductors were identified. Since these materials are expected to be stable, it should be possible to prepare them, for example by controlled halogenation of graphene (e.g., see Refs. 20,36,38), by exfoliation of the corresponding pristine materials[39] or exchange reactions starting from fluorographene, as suggested by Zboril et al.[8]

**ACKNOWLEDGMENTS**

The authors gratefully acknowledge support from the Operational Program Research and Development for Innovations - European Regional Development Fund (project CZ.1.05/2.1.00/03.0058 of the Ministry of Education, Youth and Sports of the Czech Republic), the Operational Program Education for Competitiveness - European Social Fund (project CZ.1.07/2.3.00/20.0017 of the Ministry of Education, Youth and Sports of the Czech Republic), the Barrande project (No. 7AMB12FR026) and Grant Agency of the Czech Republic (P208/11/P463 and P208/12/G016).